\documentclass[a4paper,fleqn,figfull]{cas-dc}

\usepackage[numbers]{natbib}

\begin{document}
\let\WriteBookmarks\relax
\def\floatpagepagefraction{1}
\def\textpagefraction{.001}
\shorttitle{ }
\shortauthors{M.V. Rakhlin et~al.}

\title [mode = title]{Demultiplexed Single-Photon Source with a Quantum Dot Coupled to Microresonator}


\author[1]{M.V. Rakhlin}
\cormark[1]
\ead{maximrakhlin@mail.ru}


\address[1]{Ioffe Institute, St. Petersburg, 194021, Russia}

\author[1]{A.I. Galimov}
\author[2]{I.V. Dyakonov}
\author[2]{N.N. Skryabin}
\author[1]{G.V. Klimko}
\author[1]{M.M. Kulagina}
\author[1]{Yu.M. Zadiranov}
\author[1]{S.V. Sorokin}
\author[1]{I.V. Sedova}
\author[1]{Yu.A. Guseva}
\author[1]{D.S. Berezina}
\author[1]{Yu.M. Serov}
\author[1]{N.A. Maleev}
\author[1]{A.G. Kuzmenkov}
\author[1]{S.I. Troshkov}
\author[2]{K.V. Taratorin}
\author[2]{A.K. Skalkin}
\author[2,3]{S.S. Straupe}
\author[2,4]{S.P. Kulik}
\author[1]{T.V. Shubina}
\author[1]{A.A. Toropov}

\address[2]{Quantum Technology Centre, M. V. Lomonosov Moscow State University, Moscow, 119991, Russia}
\address[3]{Russian Quantum Center, Skolkovo, Moscow, 143025, Russia}
\address[4]{South Ural State University (National Research University), Chelyabinsk, 454080, Russia}

\cortext[cor1]{Corresponding author}

\begin{abstract}
The characteristics of a single-photon emitter based on a semiconductor quantum dot, such as their indistinguishability and brightness, depend on the stability of the recombination channel, which can switch spontaneously between exciton and trion. We show that dominant recombination through neutral exciton states can be achieved by careful control of the doping profile near an epitaxial InAs/GaAs quantum dot placed in a columnar microcavity with distributed Bragg reflectors. The Hong-Ou-Mandel experiments carried out in the fabricated device demonstrate the degree of indistinguishability of 91\% of successively emitted single photons within 242 ns at an efficiency of 10\% inside a single-mode optical fiber. The achieved brightness made it possible to implement spatio-temporal demultiplexing of photons in six independent spatial modes with an in-fiber generation frequency of more than 0.1 Hz.

\end{abstract}


\maketitle

\section{Introduction}
 
The ideas of quantum information processing based on photonic circuits \cite{Flamini_2018, dale2015provable, walmsley2015quantum, kok2007linear} are currently being actively developed in the direction of their experimental implementation. The long-awaited quantum supremacy over classical computing has already been obtained for the specific problem of boson sampling  \cite{carolan2014experimental, arute2019quantum, zhong2020quantum}. The demonstrated implementation using quantum interference of 20 photons \cite{wang2019boson} indicates the fundamental feasibility of large-scale optical quantum computing schemes. A significant obstacle in this direction is the strict requirement for the parameters of the photon source, which must allow the deterministic generation of several indistinguishable photon states in parallel. The emission of two indistinguishable photons by two different quantum dots (QDs) has already been shown  \cite{zhai2021quantum}, but scaling this approach to more photons is a challenge. A more realistic way to create a multiphoton state is based on the demultiplexing of radiation from a single-photon source (SPS) generating a stream of single photons that are indistinguishable from each other \cite{hummel2019efficient, munzberg2022fast}. The temporal-to-spatial mode demultipexing is the most common scheme; in particular, it was used in  \cite{wang2019boson} to prepare a 20-photon state for the boson sampling. 

Modern SPSs intended for quantum information systems usually use the radiation of a single semiconductor QD placed either in a micropillar \cite{wang2019boson} or in a photonic waveguide \cite{hummel2019efficient}. In recent years, significant progress has been made in improving the characteristics of such sources due to the development of the microcavity design \cite{liu2019solid, tomm2021bright, wei2020bright, wang2019towards}, pumping methods \cite{wang2019towards, ding2016demand, he2019coherently}, and control of the recombination mechanism \cite{he2019coherently, somaschi2016near, tomm2021bright}. Combining the key ideas in this field, outstanding results in source brightness, single-photon purity, and indistinguishability have recently been demonstrated for an open-cavity InGaAs QD SPS \cite{tomm2021bright}.  However, the technologies used are extremely complex and the reproducible manufacture of such devices remains a challenge.

An important factor for the implementation of quantum algorithms is quantum entanglement \cite{prevedel2007photonic, gimeno2015three}. Entangled photon pairs can be generated in a biexciton radiation cascade implemented in a neutral (uncharged) QD \cite{young2007entangled}. In contrast to that, charged excitons (trions) can be an important resource for generating multiphoton cluster states \cite{Lindner}. Controlling the charge state of the QD is also important for reducing the blinking of any SPS, associated with the spontaneous recharging of the QD and successive switching between the emission of the exciton and trion states. In \cite{tomm2021bright, Hilaire}, such control is implemented by applying voltage to the p-i-n structure formed around the QD, but the implementation of this approach is quite technologically complicated.

In this work, we show that a predominantly charge-neutral state of InAs/GaAs QDs can be simply obtained by a special design of the doping profile in the heterostructure, aimed at compensating for unintentional background doping. We present the parameters of a fabricated SPS with a columnar Bragg microcavity formed from a heterostructure with QDs with such a doping profile. The high brightness of the developed SPS, along with high single-photon purity and indistinguishability, made it possible to implement the temporal-to-spatial demultiplexing of its radiation up to 6 channels.

\section{SPS design and fabrication}
The heterostructure for creating the SPS was fabricated by molecular beam epitaxy (MBE) on a GaAs:Si(001) substrate with a 500-nm-thick GaAs buffer layer. Over the buffer layer, 28(15) pairs of $\lambda$/4 Al$_{0.9}$Ga$_{0.1}$As/GaAs layers were grown, forming the lower (upper) distributed Bragg reflectors. Between them was a 266-nm-thick GaAs $\lambda$-cavity, in the center of which was an array of InAs QDs with a surface density in the range of 10$^9$-10$^{10}$ cm$^{-2}$ \cite{Galimov}, formed using the Stranski-Krastanov growth mechanism. A $p$-type doping of  $\approx$ 10$^{14}$ cm$^{-3}$ ensures that a fraction of the QDs is occupied by a single hole \cite{reference}. After measuring decay dynamics of 40 different single emission lines under resonant coherent excitation, it turned out that 80\% of the emission lines can be assigned to emission of trions. We note that of all electron-hole complexes (trion, biexciton, etc.), only the decay curve of the exciton line is modulated in time by quantum beat oscillations under $\pi$-pulse resonant excitation, induced by the fine-structure splitting of exciton levels  \cite{Ollivier}. To create neutrally charged excitons, we compensated background doping of $p$-type by introducing into the optical cavity an $n$-type Si-doped layer with an adequate thickness and the concentration of impurities $\sim$ 1$\cdot$10$^{17}$ cm$^{-3}$. This ensures that most QDs will have neutral excitons.

The regular array of micropillars with a diameter of 1 to 3.5 $\mu$m were fabricated by reactive ion-plasma etching and standard contact photolithography (365 nm) using a negative photoresist (Fig. 1a). With this approach, the spatial coincidence of the position of the QD and the center of the microcavity is a random event, as is the spectral coincidence of the QD emission line and optical resonance. The solution of these problems lies in the path of optical selection measurements for the detection of micropillars with the desired characteristics.

\begin{figure*}
\includegraphics[width=0.99\textwidth]{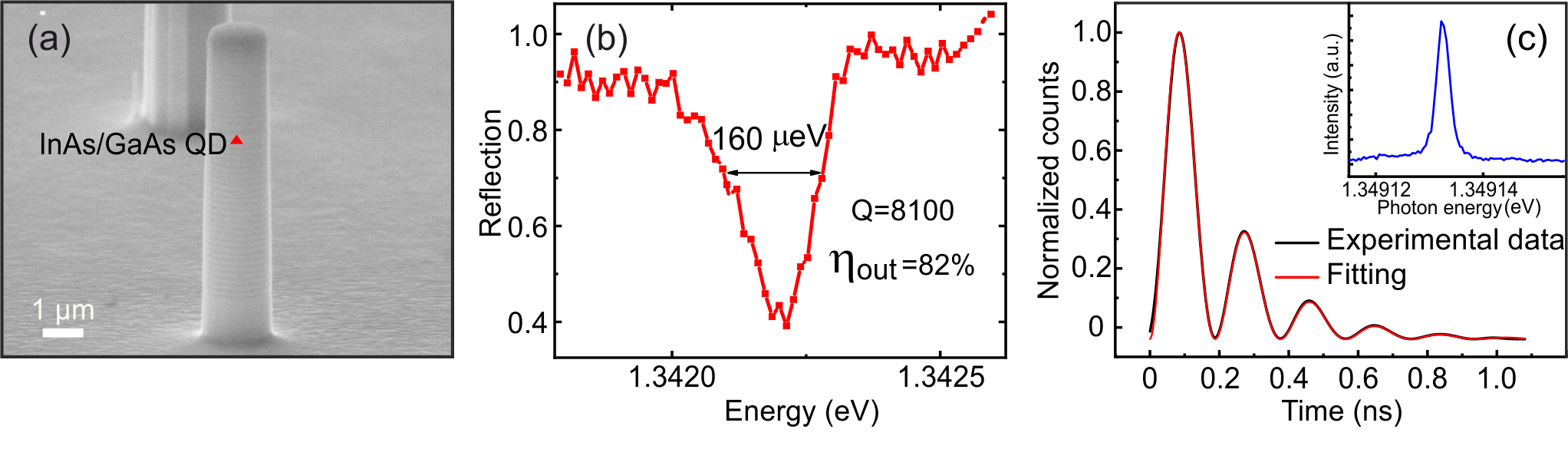}
\caption{a) Scanning electron microscopy image of the investigated micropillar with a diameter of 2 $\mu$m. b) Reflection spectrum of the micropillar  at a temperature of 8 K. (c) Deconvoluted decay curve recorded under coherent excitation by the $\pi$-pulse. Inset shows the micro-photoluminescence spectrum measured in the investigated micropillar with an InAs QD.}
\end{figure*}

\section{SPS measurements}

The efficiency (brightness) of a single-photon source ($b$) is defined as the average number of photons collected in one excitation pulse “on the first lens”, i.e. directly at the output of the photonic structure. This definition makes it possible to compare the performance of emitting devices, regardless of the efficiency of the optical scheme used to characterize them. The parameter $b$ can be expressed as \cite{Gazzano}:

\begin{equation}
\label{2f}
b=\beta\eta_{out}q_{QD},
\end{equation} 
where $\beta$  is the fraction of photons coupled into the desired cavity mode, $\eta_{out}$ is the efficiency of radiation extraction from the microcavity, $q_{QD}$ is the quantum yield associated with the QD emission line. $\eta_{out}$ is given by the ratio of the photon escape rate through the top of the cavity, $\kappa_{top}$, to the total escape rate $\kappa$:

\begin{equation}
\label{EtaOut}
\eta_{out}=\kappa_{top}/\kappa.
\end{equation} 

The parameter $\beta$  is usually defined in terms of the Purcell enhancement of the total rate of spontaneous emission \cite{Barnes}:

\begin{equation}
\beta=F_p/(F_p+1).
\end{equation} 

The Purcell factor, $F_p$, is inversely proportional to the effective mode volume $V$ \cite{Parcell}:

\begin{equation}
\label{ParcellEq}
F_p = (3/4\pi^2) (\lambda/n)^3 (Q/V), 
\end{equation} 
where $Q = \lambda /\Delta\lambda$ is the cavity quality at the wavelength $\lambda$ in vacuum and $\Delta\lambda$ is the FWHM factor of the cavity resonance; $n$ is the refractive index. 

The $\beta$ and $\eta_{out}$ parameters depend on the design of the micropillar, primarily on its diameter. According to equation ($\ref{ParcellEq}$), to enhance the emission rate, a small $V$ is needed, and the diameter must be reduced. On the other hand, if the diameter is too small, there will be less coupling between the emitter and the fundamental mode, as part of the electromagnetic field leaks out of the structure. In addition, cavity losses due to sidewall roughness will increase the value of $\kappa$ in equation ($\ref{EtaOut}$). As the diameter increases, the losses tend to saturate at the level of the planar Bragg structure. Thus, the function $\eta_{out}\beta$ is not monotonic, reaching a maximum at a diameter of about 2 $\mu$m \cite{Barnes}.

The cavity parameters for a micropillar with Bragg reflectors can be estimated by measuring the reflection spectrum from the top of the column in the resonant wavelength region. The FWHM of the dip in the spectrum gives the total cavity damping $\kappa$, and the minimum value of the reflection at the dip allows estimating $\eta_{out}$ as \cite{somaschi2016near}:

\begin{equation}
R_{min}=(1-2\eta_{out})^2.
\end{equation} 
The reflection spectrum from a micropillar 2 $\mu$m in diameter was measured using a halogen lamp (Fig. 1b). The structure under study demonstrates pretty narrow FWHM of the dip in the reflectance spectrum (160 $\mu$eV) and rather small QDs surface density. As a result, the microplillar under study contains only one emission line in the micro-photoluminescence spectrum, which is an excitonic line (see the Inset in Fig. 1c). The obtained values of the quality factor and efficiency of radiation extraction are 8100 and 82\%, respectively. 

Since the effective mode volume is not exactly known, the Purcell factor can be determined from the decrease in the lifetime of spontaneous radiative recombination in a QD placed in a resonator, $\tau$, compared to this value in the bulk material, $\tau_0$ \cite{Gerard}:

 \begin{equation}
F_p=\tau_0/\tau.
\end{equation}

\begin{figure*}
\includegraphics[width=0.99\textwidth]{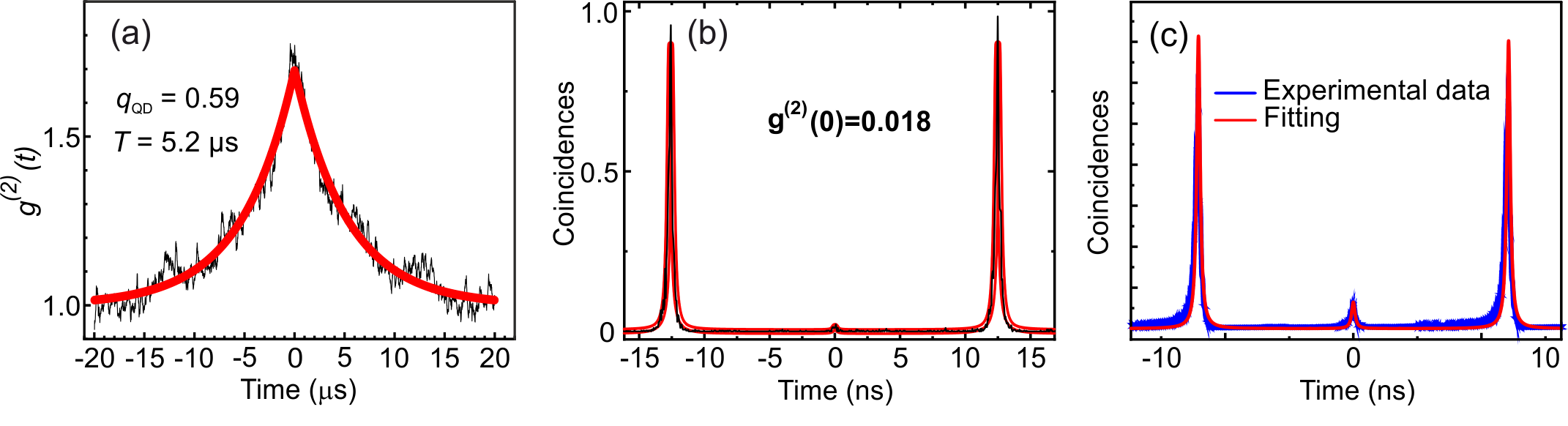}
\caption{ a) The second-order correlation function $g^{(2)}$ measured over a long time interval. The central part with the antibunching unresolved here is shown in (b). The black curve is the smoothed envelope of $g^{(2)}(\tau)$, where each point is replaced by the average value of the envelope in the range of $\pm$100 ns. The red curve represents fitting of the experimental data. b) Normalized second-order correlation function $g^{(2)}$ of single-photon radiation, measured at 8 K for the exciton line. The resulting value $g^{(2)}$(0) is 0.018. (c) Histogram of two-photon interference measured at a temperature of 8 K using the Hong-Ou-Mandel scheme with a time delay between photons of 242 ns. }

\end{figure*}

Figure 1c shows the decay dynamics of the QD exciton in the cavity, obtained with resonant coherent excitation by a $\pi$-pulse, demonstrating the emission lifetime of 184$\pm$3 ps. Since the average lifetime of an exciton state in a bulk material is about 800 ps \cite{Wang1994}, the coefficient $\beta$ can be taken to be approximately 0.8. 

The coefficient $q_{QD}$  in equation ($\ref{2f}$)  characterizes the quantum efficiency of excitation of QD radiation. In the presence of charge noise near the QD, a spontaneous switching between the charged and neutral states can occur. If the state of charge of the QD is subject to blinking, then when measuring the second-order correlation function $g^{(2)}$, bunching is observed in the extended delay region. This time dependence can be used to determine the radiation efficiency of a given exciton state. Figure 2a shows the second-order correlation function measured over a long-time interval. The black curve is the smoothed envelope of $g^{(2)}(\tau)$, each point is replaced by the average value of the envelope in the region of $\pm$100 ns. After fitting the data according to Hilaire et al.  \cite{Hilaire}, we obtained the value $q_{QD}$ = 0.59 and the average residence time of QDs in the exciton state $T$=5.2 $\mu$s. For comparison, the value $q_{QD}$ for the trion line, obtained in the reference structure without the n-type layer, , grown under the same conditions, was only 0.1, which apparently indicates a strong blinking effect in the reference structure without compensating doping. Thus, for the studied QDs, where the neutral exciton state predominates, the brightness of polarized radiation is $b$=0.32. Using a Superconducting Nanowire Single Photon Detector (SNSPD), we have registered a single photon event frequency of $6.3-6.7$ MHz, which corresponds to about 10\% efficiency inside the fiber, taking into account the detection efficiency and an 0.9 average transmission of the coupler used to connect the output fiber of the SPS and input fiber SNSPD. To our knowledge, the best end-to-end efficiency of 57\% was demonstrated by N. Tomm $et$ $al.$ using an open cavity system \cite{tomm2021bright}. However, this technology is not scalable and, in addition, requires a low-vibration cryostat with a liquid helium bath, which is extremely difficult to manufacture and operate. A result similar to that in our work was demonstrated by S.E. Thomas $et$ $al.$ reporting an overall efficacy not exceeding 11\% \cite{Thomas2021}. This value was achieved due to a complex technique of nanolithography. They also fabricated a p-i-n structure for energy adjustment and used LA-phonon-assisted excitation. So, our SPS turns out to be comparable in brightness to technologically much more complex devices. It is important to note that one can further improve the SPS performance by using recently developed promising approaches, for example, implementing linearly polarized optical pumping of an elliptic microcavity with split modes \cite{wang2019towards}.

The second-order correlation function $g^{(2)}(\tau)$, measured under strictly resonant excitation within a small time interval around 0, is shown in Fig. 2b. Fitting of the data gives the value of the correlation function $g^{(2)}(0)$ = 0.018$\pm$0.004, which is a typical value reported for best SPSs of this type \cite{somaschi2016near, tomm2021bright}.

To determine the degree of indistinguishability, experiments were carried out to study two-photon interference in the Hong-Ou-Mandel scheme for photons separated in time by 12 ns (not shown here) and 242 ns (Fig. 2c). The theoretical fitting of the data was performed according to the model proposed in \cite{Santori}. The data show that with a delay of 12 ns, the degree of indistinguishability is 0.93, and when the delay increases to 242 ns, the indistinguishability drops to 0.91. The decrease in the degree of indistinguishability with increasing time interval between photons is probably associated with spectral diffusion processes determined by charge and spin fluctuations in the material  surrounding the QD  \cite{Kuhlmann, Wang:2016, Galimov}. The relatively small contribution of such processes indicates, first of all, a rather low concentration of both defects and free charges in the structure under study.

\begin{figure*}
\includegraphics[width=0.99\textwidth]{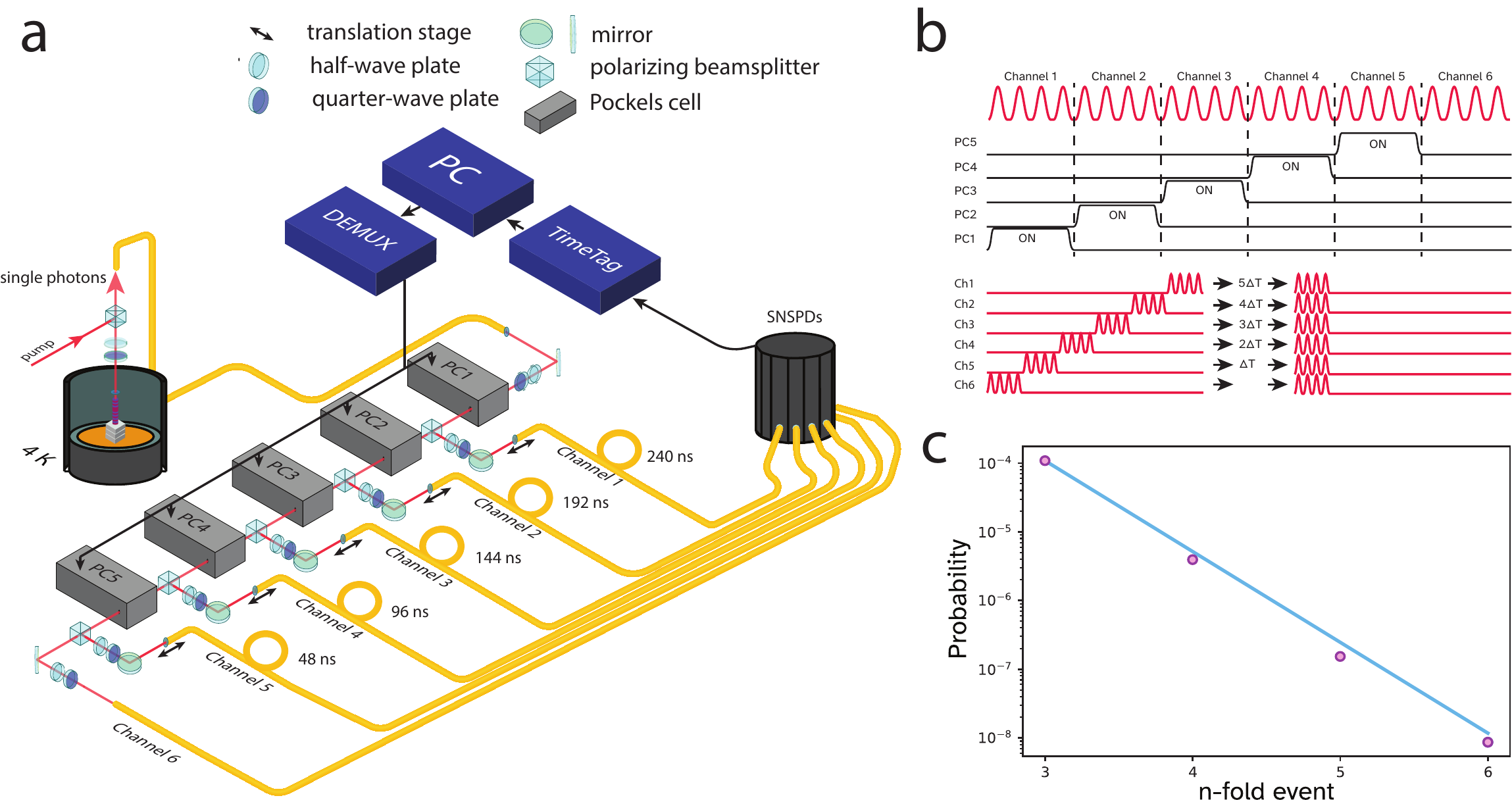}
\caption{a) Optical layout of the experimental setup. The SPS is cooled down to 4K inside an optical cryostat. The demultiplexer includes five Pockels cells (PC1, PC2, PC3, PC4, PC5) that actively pick groups of 4 pulses. The electronic FPGA-based board DEMUX receives a synchronization signal from a pulsed laser source and outputs driving voltage pulses for the Pockels cells. Each channel is coupled to a single-mode fiber coil of a certain length. Focusing aspheric lenses are mounted onto translation stages, which enables fine tuning of the output channel delays. A pair of half- and quarter-wave plates compensates for polarization rotation in the output fibers, ensuring polarization indistinguishability of photons. SNSPD block detects incoming photons and the logged data is processed by the TimeTag module. b)  An on-off sequence of each demultiplexing cycle. The delay step equals to the length of four pulse pack $\Delta T = 48.4$~ns. c) The resulting in-fiber generation probability of multiphoton events. The straight line corresponds to an exponential fit with $p^{n}$ function.}
\label{fig:demultiplexer}
\end{figure*}

\section{A multiphoton generation circuit}

The experiment with multiphoton generation was carried out in the setup described in Fig. \ref{fig:demultiplexer}a. The QD emitter was placed in an optical cryostat (Attocube attoDRY 800) at a   temperature of $3.9-4.0$~K. Spectrally filtered laser pulses generated by a femtosecond laser source (Avesta TiF-100) resonantly pumped an exciton transition in the QD in a cross-polarization configuration. Single photons at the output were collected in a single-mode fiber.

 We used a special optical circuit - a demultiplexer - to split the incoming stream of single photons into six independent channels. Figures \ref{fig:demultiplexer}a) and b) show the demultiplexer scheme and the principle of its operation, respectively. This implementation includes a set of fast (12 MHz bandwidth) electro-optical modulators. Each modulator consists of a high-voltage switch driver (BME Bergmann) and an RTP Pockels cell (Leysop) designed specifically for high-speed applications. The Pockels cells are AR coated and have a maximum transmission of 99.9\% at 925 nm. The FPGA based board is synchronized with the laser source and controls Pockels cells with a repetition rate of 1.6 MHz. Each Pockels cell is on for $1/6$ of its duty cycle and remains off the rest of the time. Pockels cells change polarization state of incoming photons and deflect them from the main axis of the optical system of the demultiplexer. The deflected photons are coupled to single-mode optical fibers using collimators with a single aspheric lens. The fiber coils with precisely tailored length synchronize the optical pulses at the outputs of the channels of the demultiplexer. The SNSPD unit registers photons coupled to a single-mode fiber at the output of the demultiplexer.

We registered multiphoton coincidence events by sending outputs of SNSPDs (detection efficiencies 0.86, 0.86, 0.87, 0.86, 0.85, 0.85) to a time-tagging electronic board TimeTag. We have registered multiphoton events occurring in the first three, four, five and six channels. Table \ref{tab:event_rates} presents results of detecting multiphoton events over an interval of 2048 seconds. We calculated the multiphoton generation probabilities in the fiber and approximated them by the exponential decay function $p^n$ (Fig. \ref{fig:demultiplexer}c) with one parameter $p$, which estimates the average value of the demultiplexing efficiency. The extracted $p=0.51$ points to unoptimized operation of the optical circuit of the demultiplexer. The main source of loss is the mode mismatch between the single- mode fiber and the incoming beam. Reflection from the fiber facet, polarizing beam splitter transmission and extinction ratio also contribute to the overall loss value in each channel.

\begin{table}
  \caption{\label{tab:event_rates}Detected multiphoton events (integration time 2048 seconds).}
  \begin{tabular*}{\tblwidth}{@{} lcccc@{} }
    \toprule
    & 3-fold & 4-fold & 5-fold & 6-fold\\
    \midrule
    Total events & 1435630&40319 & 1204 & 52\\
    Detection rate, Hz & 701.0 & 19.7 & 0.6 & 0.02\\
    \begin{minipage}{3cm}Generation rate, Hz (in fiber)\end{minipage}
    & 1494.4 & 54.2 & 2.1 & 0.1\\
    \bottomrule
  \end{tabular*}
\end{table}

\section{Conclusions}

In this work, we show the possibility of exciting predominantly neutral exciton states in self-organizing InAs/GaAs QDs placed in a Bragg resonator made in the form of a micropillar. This was done by controlling the doping profile in the active region of the structure, when the $p$-type background doping was compensated by introducing a layer of the required level of the $n$-type doping. This structure is suitable for operation as a single-photon source with the brightness of polarized radiation $b$ = 0.32. The achieved brightness made it possible to implement spatiotemporal demultiplexing of photons in six independent spatial modes with an in-fiber generation rate of more than 0.1 Hz. The results obtained show the applicability of the developed emitters of single photons as effective sources of photons for studying prototypes of a photonic quantum computer based on linear optics.

\section{Declaration of competing interest}
The authors have no conflicts of interest.

\section{Acknowledgments}
The work of M.V.R., A.I.G., G.V.K., M.M.K., Yu.M.Z., S.V.S., I.V.S., Yu.A.G., D.S.B., Yu.M.S., N.A.M., S.I.T., T.V.S. and A.A.T. was supported by Rosatom in the framework of the Roadmap for Quantum computing (Contract No. 868-1.3-15/15-2021 dated October 5, 2021 and Contract No. R2152 dated November 19, 2021). The work of I.V.D., N.N.S., K.V.T., A.K.S.  and S.S.S. was supported by Rosatom in the framework of the Roadmap for Quantum computing (Contract No. 868-1.3-15/15-2021 dated October 5, 2021 and Contract No. R2154 dated November 24, 2021).  A.I.G. and T.V.S. thank the partial support by the Russian Science Foundation, Project No. 20-42-01008 (studies of narrow exciton lines). M.V.R. also thanks the Council for Grants of the President of the Russian Federation.

\bibliographystyle{model1-num-names}
\bibliography{cas-refs}

\begin{thebibliography}{35}
\bibitem[{ref(mber)}]{reference}
\bibinfo{year}{The $p$-type doping is the unintentional background doping of
  the sample due to residuals in the {MBE} chamber}.
\end{thebibliography}

\end{document}